\documentclass{iopart}
\usepackage[dvips]{graphicx}
\def \Heff{\mathcal{H}_{\mathrm{eff}}}
\def \aver#1{\left\langle#1\right\rangle}
\def \averc#1{\left\langle#1\right\rangle_{\mathrm{conn}}}
\def \str {\mathrm{str\,}}
\def \sdet {\mathrm{sdet\,}}
\begin{document}

\title[Correlation functions of impedances and $S$-matrices]{Correlation
functions of impedance and scattering matrix elements in chaotic absorbing
cavities}
\author{D. V. Savin$^1$, Y. V. Fyodorov$^2$, and H.-J. Sommers$^1$}
\address{$^1$\,Fachbereich Physik, Universit\"at Duisburg-Essen, 45117 Essen, Germany}
\address{$^2$\,School of Mathematical Sciences, University of
Nottingham, Nottingham NG7 2RD, UK}

\begin{abstract}
Wave scattering in chaotic systems with a uniform energy loss (absorption)
is considered. Within the random matrix approach we calculate exactly the
energy correlation functions of different matrix elements of impedance or
scattering matrices for systems with preserved or broken time-reversal
symmetry. The obtained results are valid at any number of arbitrary open
scattering channels and arbitrary absorption. Elastic enhancement factors
(defined through the ratio of the corresponding variance in reflection to
that in transmission) are also discussed.
\end{abstract}


\submitto{Acta Physica Polonica A \qquad 22 June 2005\\ (Proceedings of the
2nd Workshop on Quantum Chaos and Localization Phenomena, \\ May 19--22, 2005, Warsaw)} %

\section{Introduction}

Propagation of electromagnetic or ultrasonic waves in billiards
\cite{Stoeckmann}, scattering of light in random media and transport of
electrons through quantum dots \cite{Beenakker1997,Alhassid2000} share at
least one feature in common: In all these situations one deals with an open
wave-chaotic system studied by means of a scattering experiment, see Fig.~1
for an illustration. Here, we have a typical transport problem where the
fundamental object of interest is the scattering matrix $S$, which relates
linearly the amplitudes of incoming and outgoing fluxes. However, under real
laboratory conditions there is a number of different sources which cause
that a part of the flux gets irreversibly lost or dissolved in the
environment. As a result, we encounter absorption and have to handle the
$S$-matrix, which is no longer unitary. Statistics of different scattering
observables in the presence of absorption are nowadays under intense
experimental and theoretical studies. One should mention, in particular,
experiments on energy correlations of the $S$-matrix
\cite{Doron1990,Lewenkopf1992i} and total cross-sections
\cite{Schaefer2003}, distributions of reflection
\cite{Doron1990,Mendez-Sanchez2003} and transmission \cite{Schanze2005} as
well as that of the complete $S$ matrix \cite{Kuhl2005} in microwave
cavities, properties of resonance widths \cite{Barthelemy2005a} in such
systems at room temperatures, dissipation of ultrasonic energy in
elastodynamic billiards \cite{Lobkis2003}, fluctuations in microwave
networks \cite{Hul2004} (see also references in these papers).
Theoretically, statistics of reflection, delay times and related quantities
were considered first in the strong \cite{Kogan2000} or weak
\cite{Beenakker2001} absorption limits at perfect coupling, and very
recently  at arbitrary absorption and coupling
\cite{Savin2003i,Fyodorov2003i,Savin2004,Rozhkov2003,Rozhkov2004}.

Another insight to the same problem comes by considering it not from the
``outside'', but rather from ``inside''. Then the impedance relating
linearly a voltage to a current turns out to be the prime object of interest
\cite{Hemmady2005,Hemmady2005i}, see Fig.~1. It turns out that after proper taking
into account of the wave nature of the current \cite{Zheng2004a,Zheng2004b}, the
cavity impedance becomes an electromagnetic analogue of Wigner's reaction
($R$) matrix of the scattering theory. This can be understood qualitatively
through the well-known equivalence of the two-dimensional Maxwell equations
to the Schr\"odinger equation, the role of the wave function being played by
the field (the voltage in our case). Then the definition of the impedance
becomes formally similar to the definition of the $R$-matrix (which relates
linearly the normal derivative of the wave function to the wave function
itself on the boundary). The impedance is, therefore, related to the local
Green function of the closed cavity and fluctuates strongly due to chaotic
internal dynamics.

The imaginary part of the local Green function (which is proportional to the
real part of the impedance) is known in the context of mesoscopics as the
local density of states and has a long story of study, see \cite{Mirlin2000}
for a recent review. Actually, a closely related quantity in the context of
spectra of complex atoms and molecules has the meaning of the total
cross-section of indirect photoabsorption, see e.g. \cite{Fyodorov1998i}. As
to the real part, it seems to have no direct physical meaning in mesoscopics
while it has the meaning of reactance in electromagnetics, where both real
and imaginary parts are experimentally studied. Very recently an approach
\cite{Fyodorov2004ii,Savin2005} has been developed by us which allows one to
study the (joint) distribution function of these real and imaginary parts at
arbitrary absorption and to relate it to the reflection distribution, thus
linking somewhat complementary experiments \cite{Kuhl2005} and
\cite{Hemmady2005} together.
\begin{figure}
\mbox{\ }\hfill\includegraphics[width=0.8\textwidth]{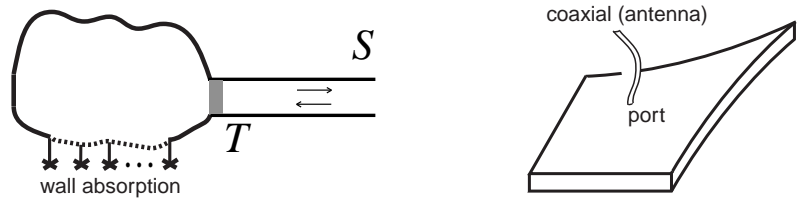}\hfill
\caption{A sketch of a typical experimental setup with microwave billiards.
A flat chaotic cavity is feeded with microwaves through an attached coaxial
cable (i.e. a scattering channel). On average, $1-T$ part of the incoming
flux, where $T\leq1$ is the so-called transmission coefficient, is reflected
back directly from the cable-cavity interface (port) without exciting
long-lived resonances in the cavity. If the cavity is thin enough then only
a transverse electric wave can propagate inside. The electric field has only
a vertical component, which is uniform in vertical direction and distributed
nontrivially in the plane. Therefore, there is a voltage between plates as
well as a current due to the in-plane magnetic field. The impedance is a
quantity which relates linearly the port voltage to the port current.
Fluctuations of eigenmodes and eigenfrequencies result in fluctuations of
the impedance or $S$-matrix, as the driving frequency or port position is
changed.}
\end{figure}

Due to a strong resonance energy dependence the impedance and  $S$-matrix as
well as any  scattering observable exhibit strong fluctuations over a
smooth regular background as the scattering energy (or another external
parameter) is varied. These two variations occurring on different energy
scales are usually decomposed into a mean and a fluctuating part by means of
the spectral or (assumed to be equivalent) ensemble average
$\langle\cdots\rangle$. In this paper we consider statistics as determined
by a two-point correlation function of the fluctuating parts (also called a
``connected'' correlator): $\averc{AB}=\aver{AB}-\aver{A}\aver{B}$. We
restrict ourselves below to the cases of preserved and broken time-reversal
symmetry (TRS).

\section{Scattering, RMT and absorption}

The resonance energy dependence of observables becomes explicit in the
well-known Hamiltonian approach to quantum scattering, which was developed
first in the context of  nuclear physics
\cite{Mahaux,Verbaarschot1985,Sokolov1989} and adopted later for the needs
of mesoscopic physics, see e.g.
\cite{Alhassid2000,Lewenkopf1991,Fyodorov1997}. This framework is adequate
to take finite absorption into account as well. We have the following
relation between the resonance part of the scattering matrix and Wigner's
reaction matrix:
\begin{equation}\label{S}
S(E) = \frac{1-\rmi K(E)}{1+\rmi K(E)}\,, \qquad
K(E)=\case12V^{\dag}(E-H)^{-1}V\,.
\end{equation}
The Hamiltonian $H$ of the \emph{closed} system gives rise to $N$ levels
(eigenfrequencies) which are coupled to $M$ continuum channels via the
$N{\times}M$ matrix $V$ of coupling amplitudes $V^c_n$ ($n=1\ldots N$,
$c=1\ldots M$). Performing for $S$ a Taylor series expansion in $K$ and
regrouping the terms, one comes to another well-known expression for the $S$
matrix
\begin{equation}\label{Heff}
S(E) = 1-\rmi V^{\dag}\frac{1}{E-\Heff}V\,, \qquad
\Heff=H-\case\rmi2VV^{\dag}\,.
\end{equation}
in terms of the effective Hamiltonian $\Heff$ of the \emph{open} system,
which is non-Hermitian contrary to the Hermitian $H$. The factorized
structure of the anti-Hermitian part ensures the unitarity of $S(E)$ at real
values of $E$. In a resonance approximation of the energy-independent
amplitudes the complex eigenvalues $\mathcal{E}_n=E_n-\case\rmi2\Gamma_n$ of
$\Heff$ are the only singularities of the $S$ matrix in the complex energy
plane. As required by causality \cite{Nussenzveig}, they are located in the
lower half plane and correspond to the long-lived resonance states, with
energies $E_n$ and escape widths $\Gamma_n>0$, which are formed on the intermediate
stage of a scattering process.

To mimic chaotic nature of the intrinsic motion we adopt, as usual, the
random matrix theory (RMT) \cite{Beenakker1997,Mehta2,Guhr1998} and replace
the actual Hamiltonian with a random Hermitian matrix $H$. It turns out that
spectral fluctuations possess a large degree of universality in the limit
$N{\to}\infty$: being expressed (``unfolding'') in units of the mean level
spacing $\Delta$ they become independent of microscopic details (i.e. a
particular form of the distribution of $H$) and get uniformly distributed
throughout the whole spectrum \cite{Guhr1998}. That amounts usually to
considering \emph{local} fluctuations at the center of the spectrum ($E=0$)
and to restricting ourselves to the simplest case of Gaussian ensembles. On
has the Gaussian Orthogonal Ensemble (GOE, the Dyson's symmetry index
$\beta=1$ and $H$ symmetric) for chaotic systems with preserved
time-reversal symmetry (TRS) and the Gaussian Unitary  Ensemble (GUE,
$\beta=2$ and $H$ Hermitian) for those with fully broken TRS. For similar
reasons the approach is independent of particular statistical assumptions on
coupling amplitudes $V^{a}_{n}$ (as long as $M\ll N$
\cite{Lehmann1995a,Lehmann1995b}), which may be chosen as fixed
\cite{Verbaarschot1985} or random \cite{Sokolov1989} variables. They enter
final expressions only by means of $M$ transmission coefficients (also
so-called sticking probabilities)
\begin{equation}\label{T}
T_c\equiv1-|\overline{S^{cc}}|^2=\frac{4\kappa_c}{(1+\kappa_c)^2}\,, \qquad
\kappa_c=\frac{\pi\|V^c\|^2}{2N\Delta}\,.
\end{equation}
where $\overline{S^{cc}}$ stands for the average (or ``optical'') $S$
matrix. They are assumed to be input parameters of the theory. $T_c\ll1$ or
$T_c=1$ corresponds to an almost closed or perfectly open channel ``c'',
respectively.

Absorption is usually seen as a dissipation process, which evolves
exponentially in time. Strictly speaking, different spectral components of
the field have different dissipation rates. However, this rather weak energy
dependence can easily be neglected as long as local fluctuations on much
finer energy scale $\sim\Delta$ are considered. As a result, all the
resonances acquire  additionally to their escape widths one and the same
absorption width $\Gamma>0$. The dimensionless parameter
$\gamma\equiv2\pi\Gamma/\Delta$ characterizes then the absorption strength,
with $\gamma\ll1$ or $\gamma\gg1$ corresponding to the weak or strong
absorption limit, respectively. (Microscopically, it can be modelled by
means of a huge number of weakly open parasitic channels
\cite{Lewenkopf1992i,Brouwer1997ii} or by additional coupling to very
complicated background with almost continuous spectrum \cite{Savin2003i},
see also \cite{Sokolov2005}.)

Treating $\Gamma$ phenomenologically, one sees that such a \emph{uniform}
absorption can equivalently be taken into account by a purely imaginary
shift of the scattering energy $E\to E+\case\rmi2\Gamma\equiv E_{\gamma}$,
so that the $S$-matrix $S_{\gamma}(E)\equiv S(E_{\gamma})$ becomes
subunitary. The reflection matrix $S_{\gamma}^{\dag}S_{\gamma}<1$ provides
then a natural measure of the mismatch between incoming and outgoing fluxes
\cite{Beenakker2001,Savin2003i}. At last but not least, the matrix $Z\equiv\rmi
K(E_{\gamma})$ has the meaning of the normalized cavity impedance in such a
setting, see \cite{Zheng2004a,Zheng2004b} for further
details.

\section{Correlation functions}

\subsection{Impedance}%
Let us consider first the simplest case of the impedance when the problem
can be fully reduced to that of spectral correlations determined by the
two-point cluster function $Y_{2,\beta}(\omega)=\delta(\omega)
-\Delta^2\aver{\rho(E_1)\rho(E_2)}_{\mathrm{conn}}$, where
$\omega=(E_2-E_1)/\Delta$ and $\rho(E)$ being the spectral density. It is
easy to find $\aver{Z^{ab}}=\kappa_{a}\delta^{ab}$ for the mean impedance at
$E=0$. To calculate the energy correlation function
\begin{equation}\label{Zcorr}
C^{abcd}_Z(\omega)\equiv\averc{Z^{ab*}(E_1) Z^{cd}(E_2)}
\end{equation}
it is instructive to write
$Z^{ab}(E)=\case\rmi2\sum_{n}\frac{v^{a*}_nv^b_n}{E-E_{n}+\rmi\Gamma/2}$ in
the eigenbasis of the closed system. The rotation that diagonalizes the
random $H$ transforms the (fixed) coupling amplitudes $V^{a}_n$ to gaussian
distributed random coupling amplitudes $v^a_n$ with the zero mean and the
second moment
$\aver{v^{a*}_nv^{b}_m}=(2\kappa_a\Delta/\pi)\delta^{ab}\delta_{nm}$. In
such a representation (\ref{Zcorr}) acquires the following form:
\begin{equation}\label{Zcorr1}
\fl
C^{abcd}_Z(\omega)=\sum_{n,m}\frac{1}{4}\aver{v^{a}_nv^{b*}_nv^{c*}_mv^{d}_m}
\averc{ \frac{1}{E_1-E_n-\case\rmi2\Gamma}
\frac{1}{E_2-E_m+\case\rmi2\Gamma} }
\end{equation}
so that averaging over coupling amplitudes (i.e. eigenfunctions) and that
over the spectrum can be done independently. The gaussian statistics of $v$
results in
\begin{equation}\label{vvvv}
\fl \case14(\case{\pi}{\Delta})^2 \aver{v^{a}_nv^{b*}_nv^{c*}_mv^{d}_m}=
\kappa_a\kappa_c\delta^{ab}\delta^{cd} +
\kappa_a\kappa_b(\delta^{ac}\delta^{bd} +
\delta_{1\beta}\delta^{ad}\delta^{bc})\delta_{nm}
\end{equation}
where $\delta_{1\beta}$ term accounts for the presence of TRS, when all
$v^a_n$ are real and $Z$ is symmetric. It is useful then to represent the
spectral correlator in the form of the Fourier integral
$\int_{0}^{\infty}\rmd t_1\int_{0}^{\infty}\rmd t_2 e^{-\Gamma(t_1+t_2)/2}
e^{\rmi E(t_2-t_1)} e^{\rmi(E_2-E_1)(t_1+t_2)/2}
\averc{e^{\rmi(E_nt_1-E_mt_2)}}$. Due to the uniformity of local
fluctuations in the bulk of the spectrum, one can integrate additionally
over the position $E$ of the mean energy: $\int\!\frac{\rmd
E}{N\Delta}e^{\rmi E(t_2-t_1)}=\case1N\delta(\frac{t_2-t_1}{t_H})$, where
$t_{H}\equiv2\pi/\Delta$ is the Heisenberg time. From the known RMT spectral
fluctuations one also has
$(1-N)\averc{e^{\rmi(E_n-E_m)t}}=b_{2,\beta}(t/t_H)$ for $n{\ne}m$, where
$b_{2,\beta}(\tau)$ is the spectral form factor defined through the Fourier
transform of $Y_{2,\beta}(\omega)$ \cite{Mehta2,Guhr1998}:
\numparts%
\begin{eqnarray}
\fl \label{b_21}%
b_{2,\beta=1}(\tau)=[1-2\tau+\tau\log(1+2\tau)]\Theta(1-\tau)+
[\tau\log(\case{2\tau+1}{2\tau-1})-1]\Theta(\tau-1)
\\
\fl \label{b_22}%
b_{2,\beta=2}(\tau)= (1-\tau)\Theta(1-\tau)
\end{eqnarray}
\endnumparts%
at $\tau>0$ and $b_{2,\beta}(-\tau)=b_{2,\beta}(\tau)$, so that
$Y_{2,\beta}(\omega)=\int_{-\infty}^{\infty}\!\rmd\tau
e^{2\pi\rmi\omega\tau}b_{2,\beta}(\tau)$. Combining all these results
together and measuring the time in units of the Heisenberg time
($\tau=t/t_H$), we arrive finally at
\numparts%
\begin{eqnarray}
\fl \label{ZcorrFTa}%
C^{abcd}_Z(\omega)=\int_0^{\infty}\!\!\!\rmd\tau e^{2\pi\rmi\omega\tau}
C^{abcd}_Z(\tau)
\\
\fl \label{ZcorrFTb}%
C^{abcd}_Z(\tau)=4\,e^{-\gamma\tau}\,[\kappa_a\kappa_c(1-b_{2,\beta}(\tau))
\delta^{ab}\delta^{cd} + \kappa_a\kappa_b(\delta^{ac}\delta^{bd} +
\delta_{1\beta}\delta^{ad}\delta^{bc}) ]\,.
\end{eqnarray}
\endnumparts%
Similar in spirit calculations were done earlier a context of
reverberation in complex structures in \cite{Davy1987,Lobkis2000} and
in a context of chaotic photodissociation in
\cite{Alhassid1992,Alhassid1998}.

The form factor (\ref{ZcorrFTb}) is simply related  to that of $K$ matrix
elements at zero absorption as
$C^{abcd}_Z(\tau)=e^{-\gamma\tau}C^{abcd}_K(\tau)$. Such a relationship
between the corresponding form factors with and without absorption is
generally valid for any correlation function which may be reduced to the
two-point correlator of resolvents (see \cite{Schaefer2003} and below, e.g.,
for the case of the $S$ matrix). This can be easily understood as the
result of the analytic continuation $2\pi\omega\to2\pi\omega+i\gamma$ of the
energy difference $\omega$ when absorption is switched on (see the previous
section).

\begin{figure}
\mbox{\ }\hfill\includegraphics[width=0.85\textwidth]{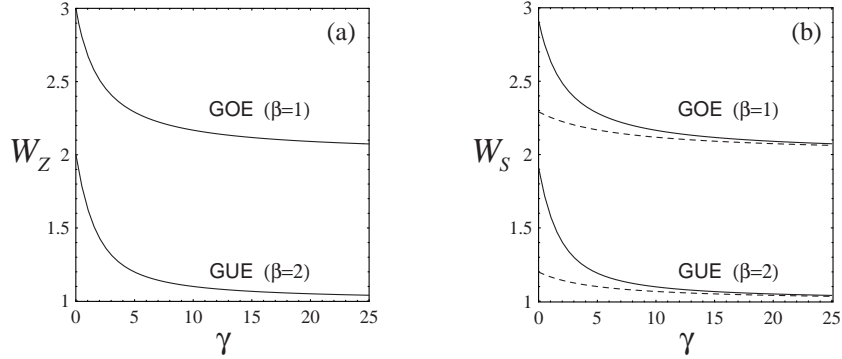}
\caption{The impedance (a) and $S$-matrix (b) enhancement factors for
chaotic systems with preserved ($\beta=1$) or broken ($\beta=2$)
time-reversal symmetry as functions of the absorption strength $\gamma$. The
case (b) corresponds to the many-channel limit with isolated
($\sum_cT_c=0.2$, solid lines) or overlapping resonances ($\sum_cT_c=5$,
dashed lines).}
\end{figure}
The obtained expressions describe a decorrelation process of the $Z$ matrix
elements as the energy difference grows, generally,
$C_Z(\omega{\to}\infty)\to0$. At $\omega=0$, (\ref{ZcorrFTa}) provides us
with impedance variances $C_Z^{abab}(0)=\mathrm{var}(Z^{ab})\equiv
\langle|Z^{ab}|^2\rangle-|\langle{Z^{ab}}\rangle|^2$, which were recently
studied in \cite{Zheng2005} (see also \cite{Hemmady2005ii}).
In analogy with the so-called elastic
enhancement factor considered frequently in nuclear physics
\cite{Verbaarschot1986}, one can define the following ratio of variances in
reflection ($a{=}b$) to that in transmission ($a{\neq}b$):
\begin{equation}\label{W_Z}
W_{Z,\beta} \equiv \frac{\sqrt{\mathrm{var}(Z^{aa})\mathrm{var}(Z^{bb})} }{
\mathrm{var}(Z^{ab})} = 2+\delta_{1\beta} - \int_{0}^{\infty}\!\!\rmd s\,
e^{-s} b_{2,\beta}(\case{s}{\gamma})
\end{equation}
where the second equality follows easily from (\ref{ZcorrFTb}) (note that
the coupling constants $\kappa_{a,b}$ are mutually cancelled here). Making
use of $b_{2,\beta}(\infty){=}0$ and $b_{2,\beta}(0){=}1$, one can readily
find $W_{Z,\beta}$ in the limiting cases of weak or strong absorption as:
\begin{equation}\label{W_Zlim}
W_{Z,\beta} =\cases{2+\delta_{1\beta} & at $\gamma\ll1$\\
1+\delta_{1\beta} & at $\gamma\gg1$\\} \,.
\end{equation}
$W_{Z,\beta}$ decays monotonically as absorption grows, see Fig.~1a. In the
case of unitary symmetry, (\ref{b_22}) and (\ref{W_Z}) yield explicitly
$W_{Z,2}=1+\case1\gamma(1-e^{-\gamma})$ in agreement with \cite{Zheng2005}.
It is hardly possible to get a simple explicit expression at finite $\gamma$
in the case of orthogonal symmetry. However, a reasonable approximation can
be found if one notices that the integration in (\ref{W_Z}) is determined
mainly by the region $s\leq1$, so that one can approximate
$b_{2,1}(s)\approx(1-2s+2s^2)\Theta(1-s)$ through its Taylor expansion.
Performing the integration, one arrives at
$W_{Z,1}\approx3-\gamma^{-2}[(4+\gamma^2)(1-e^{-\gamma})
-2\gamma(1+e^{-\gamma})]$, which turns out to be a good approximation to the
exact answer at moderate absorption (deviations are seen numerically only at
$\gamma\sim1$).

\subsection{$S$-matrix elements}

The energy correlation function of the scattering matrix elements
\begin{equation}\label{Scorr}
C^{abcd}_S(\omega)\equiv\averc{S_{\gamma}^{ab*}(E_1)S_{\gamma}^{cd}(E_2)}
=\int_0^{\infty}\!\!\rmd\tau e^{2\pi\rmi\omega\tau} C^{abcd}_S(\tau)
\end{equation}
is a much more complicated object for an analytical treatment as
(\ref{Zcorr}). The reason becomes clearer if one considers again the pole
representation of the $S$ matrix which follows from (\ref{Heff}):
$S^{ab}(E)=\delta^{ab}-\rmi\sum{w^{a}_n\tilde{w}^{b}_n}/(E-\mathcal{E}_n)$.
Due to a unitarity constraint imposed on $S$ (at real $E$), the residues and
complex energies get mutually correlated \cite{Sokolov1989} with a generally
unknown joint distribution. The separation like (\ref{Zcorr1}) into a
``coupling'' and ``spectral'' average in no longer possible and can be done
only by involving some approximations \cite{Gorin2002}. The powerful
supersymmetry method \cite{Verbaarschot1985,Efetov1983} turns out to be an
appropriate technique to perform the statistical average in this case. In
their seminal paper \cite{Verbaarschot1985}, Verbaarschot, Weidenm\"uller
and Zirnbauer performed the exact calculation of (\ref{Scorr}) at arbitrary
transmission coefficients (and zero absorption) in the case of orthogonal
symmetry. This finding was later adopted \cite{Schaefer2003} to include
absorption. The corresponding exact result for unitary symmetry is still
lacking in the literature (see, however, \cite{Pluhar1995} concerning the
$S$-matrix variance in the GOE-GUE crossover at perfect coupling) and will
be presented below.

The calculation proceeds alone the same line as in \cite{Verbaarschot1985},
we indicate only essential differences.  As usual, the representation of
resolvents and thus (\ref{Scorr}) in the form of gaussian integrals over
auxiliary ``supervectors'' consisting of both commuting and anticommuting
(Grassmann) variables allows one to perform statistical averaging exactly.
In the limit $N\to\infty$, the rest integration over the auxiliary field can
be done in the saddle-point approximation.  The final expression for both
the correlator and its form factor (\ref{Scorr}) can be equally represented
as follows:
\begin{eqnarray}\label{ScorrFT}
\fl C^{abcd}_S= \delta^{ab}\delta^{cd}T_aT_c\sqrt{(1-T_a)(1-T_c)} J_{ac} +
(\delta^{ac}\delta^{bd}+\delta_{1\beta}\delta^{ad}\delta^{bc})T_aT_bP_{ab}\,.
\end{eqnarray}
Here, the $\delta_{1\beta}$ term accounts trivially for the symmetry
property $S^{ab}=S^{ba}$ in the presence of TRS. $J_{ac}$ and $P_{ab}$
defined below are some functions (of the energy difference $\omega$ or the
time $\tau$), which depend also on TRS, coupling and absorption but already
in a nontrivial way. As a result, the elastic enhancement factor
$W_{S,\beta}\equiv\sqrt{\mathrm{var}(S^{aa})\mathrm{var}(S^{bb})}/\mathrm{var}(S^{ab})$
is generally a complicated function of all these parameters, in contrast to
(\ref{W_Z}). In the particular case of perfect coupling, all $T_{c}=1$, one
has obviously from (\ref{ScorrFT}) that $W_{S,\beta}=1+\delta_{1\beta}$ at
any absorption strength.

The saddle-point integration turns out to have a nontrivial saddle-point
manifold \cite{Efetov1983} over which one needs to integrate exactly. This
task can be accomplished by making use of the ``angular'' parametrization
\cite{Verbaarschot1985} of the manifold in terms of the
$\frac{4}{\beta}{\times}\frac{4}{\beta}$ supermatrices $t_{12}$ and $t_{21}$.
We consider first real $\omega$ (no absorption). Then the
functions $J_{ac}$ and $P_{ab}$  have in the energy domain the following
representation:
\numparts %
\begin{eqnarray}
\fl \label{J}%
J_{ac}(\omega) = \frac{\beta^2}{16} \aver{
\str\Bigl[\frac{t_{12}t_{21}}{1+T_at_{12}t_{21}}k\Bigr]\,
\str\Bigl[\frac{t_{21}t_{12}}{1+T_ct_{21}t_{12}}k\Bigr]
\mathcal{F}^{(\beta)}_M}_{\mu}
\\
\fl \label{P}%
P_{ab}(\omega) = \frac{\beta}{16} \aver{
\str\Bigl[t_{21}\frac{(1+t_{12}t_{21})^{1/2}}{1+T_at_{12}t_{21}}
kt_{12}\frac{(1+t_{21}t_{12})^{1/2}}{1+T_bt_{21}t_{12}}k\Bigr]
\mathcal{F}^{(\beta)}_M}_{\mu}
\end{eqnarray}
\endnumparts
that is completely in a parallel with \cite{Verbaarschot1985} (the diagonal
matrix $k=1$ ($-1$) in the subspace of commuting (anticommuting) variables).
This result has the form of an expectation value
$\langle(\cdots)\rangle_{\mu}\equiv \int\!
\rmd\mu_{(\beta)}(\cdots)e^{\rmi\mathcal{L_{\beta}}(\omega)}$ in the field
theory (nonlinear ``zero-dimensional'' supersymmetric $\sigma$-model)
characterized by the Lagrangian
$\mathcal{L}_{\beta}(\omega)=\beta\pi\omega\,\str(t_{12}t_{21})$. The
so-called channel factor
$\mathcal{F}^{(\beta)}_M=\prod_{c=1}^{M}\sdet(1+T_ct_{12}t_{21})^{-\beta/2}$
accounts for system openness. We refer the reader to
\cite{Verbaarschot1985,Efetov} for a definition of the supertrace and
superdeterminant as well as for a general discussion of the superalgebra. An
explicit parametrization of matrices $t_{12},t_{21}$ and the integration
measure $\rmd\mu_{(\beta)}$ over them depend on the symmetry case
considered; it can be found in \cite{Verbaarschot1985} for $\beta=1$ and in
\cite{Zuk1994,Fyodorov1995} for $\beta=2$. Essential is that the final
expressions are determined only by real ``eigenvalues'' $\mu_0$ and
$\mu_{1,2}$ of the angular matrices. Finally, one can cast resulting
expressions as follows
\numparts %
\begin{eqnarray}
\fl J_{ac}(\omega) = \textstyle \aver{
\bigl(\frac{\mu_1}{1+T_a\mu_1}+\frac{\mu_2}{1+T_a\mu_2}+\frac{\mu_0}{1-T_a\mu_0}\bigr)
\bigl(\frac{\mu_1}{1+T_c\mu_1}+\frac{\mu_2}{1+T_c\mu_2}+\frac{\mu_0}{1-T_c\mu_0}\bigr)
\mathcal{F}_M}_{\mu}
\\ \textstyle
\fl P_{ab}(\omega) = \textstyle \aver{ \bigl(
\frac{\mu_1(1+\mu_1)}{(1+T_a\mu_1)(1+T_b\mu_1)}
+\frac{\mu_2(1+\mu_2)}{(1+T_a\mu_2)(1+T_b\mu_2)}
+\frac{\mu_0(1-\mu_0)}{(1-T_a\mu_0)(1-T_b\mu_0)} \bigr) \mathcal{F}_M}_{\mu}
\end{eqnarray}
\endnumparts
with
$\mathcal{F}^{}_{M}=\prod_c[\frac{(1-T_c\mu_0)^2}{(1+T_c\mu_1)(1+T_c\mu_2)}]^{1/2}$
in the $\beta=1$ case of orthogonal symmetry \cite{Verbaarschot1985}, and
\numparts %
\begin{eqnarray}
J_{ac}(\omega) = \textstyle \aver{
\bigl(\frac{\mu_1}{1+T_a\mu_1}+\frac{\mu_0}{1-T_a\mu_0}\bigr)
\bigl(\frac{\mu_1}{1+T_c\mu_1}+\frac{\mu_0}{1-T_c\mu_0}\bigr)
\mathcal{F}_M}_{\mu}
\\ \textstyle
P_{ab}(\omega) = \textstyle \aver{ \bigl(
 \frac{\mu_1(1+\mu_1)}{(1+T_a\mu_1)(1+T_b\mu_1)}
+\frac{\mu_0(1-\mu_0)}{(1-T_a\mu_0)(1-T_b\mu_0)} \bigr) \mathcal{F}_M}_{\mu}
\end{eqnarray}
\endnumparts
with $\mathcal{F}_{M}=\prod_c\frac{1-T_c\mu_0}{1+T_c\mu_1}$ in the $\beta=2$
case of unitary symmetry. Here, the corresponding integration
$\langle(\cdots)\rangle_{\mu}$ is to be understood explicitly  for these two
respective cases as
\begin{equation}
\fl \frac{1}{8} \int_{0}^{\infty}\!\rmd\mu_1 \!\int_{0}^{\infty}\!\rmd\mu_2
\!\int_{0}^{1}\!\rmd\mu_0 \frac{(1-\mu_0)\mu_0
|\mu_1-\mu_2|\,e^{\rmi\pi\omega(\mu_1+\mu_2+2\mu_0)} }{
[(1+\mu_1)\mu_1(1+\mu_2)\mu_2]^{1/2} (\mu_0+\mu_1)^2 (\mu_0+\mu_2)^2 }
\left(\ldots \right)
\end{equation}
  and
\begin{equation}
\int_{0}^{\infty}\!\rmd\mu_1\int_{0}^{1}\!\rmd\mu_0
(\mu_1+\mu_0)^{-2}\,e^{\rmi2\pi\omega(\mu_1+\mu_0)} \left(\ldots \right)\,.
\end{equation}
In the important particular case of the single open channel (elastic
scattering), the general expression for the $\beta=2$ case simplifies
further to
\begin{equation}
\fl \averc{S^*(E_1)S(E_2)}=T^2\int_{0}^{\infty}\!\rmd\mu_1\int_{0}^{1}
\!\frac{\rmd\mu_0}{\mu_1+\mu_0}
\frac{1+(2-T)\mu_1}{(1+T\mu_1)^3}\,e^{\rmi2\pi\omega(\mu_1+\mu_0)}\,.
\end{equation}
Finally, putting above $\omega\to\omega+\rmi\gamma/2\pi$ accounts for the
finite absorption strength $\gamma$.

To consider (\ref{ScorrFT}) in the time domain, i.e. the form factor
$C^{abcd}_S(\tau)$, we notice that the variable
$\tau=\case12(\mu_1+\mu_2+2\mu_0)$ for $\beta=1$ or $\tau=\mu_1+\mu_0$ for
$\beta=2$ plays the role of the dimensionless time. The corresponding
expressions for $P_{ab}(\tau)$ and $J_{ac}(\tau)$ can be investigated using
the methods developed in \cite{Verbaarschot1986,Gorin2002,Dittes2000}. For
orthogonal symmetry it was done in \cite{Schaefer2003}, where the overall
decaying factor $e^{-\gamma\tau}$ due to absorption was also confirmed by
comparison to the experimental result for the form factor measured in
microwave cavities. It is useful for the qualitative description to note
that $P_{ab}(\tau)$ and $2J_{ac}(\tau)$ are quite similar to the ``norm
leakage'' decay function \cite{Savin1997} and the form factor of the
Wigner's time delays \cite{Lehmann1995b}, respectively (they would coincide
exactly at $\gamma=0$, if we put $T_{a,b,c}=0$ appearing explicitly in
denominators of (\ref{J}) and thereafter). Then one can follow analysis
performed in these papers, see also \cite{Dittes2000}, to find qualitatively
$P_{ab}(\tau)\sim e^{-\gamma\tau}(1+\case2\beta T_a\tau)^{-1} (1+\case2\beta
T_b\tau)^{-1}\prod_{c}(1+\case2\beta T_c\tau)^{-\beta/2}$ and
$J_{ab}(\tau)\sim[1-b_{2,\beta}(\tau)]P_{ab}(\tau)$. One has
$P_{ab}(\tau)\approx e^{-\gamma\tau}$ and
$J_{ab}(\tau)\approx(2\tau/\beta)e^{-\gamma\tau}$ as exact asymptotic at
small times \cite{Gorin2002}, they both being $\sim
e^{-\gamma\tau}\tau^{-M\beta/2-2}$ at large times.

Such a power law  is characteristic for open systems
\cite{Lewenkopf1991,Dittes2000,Savin1997}. Physically, it results from width
fluctuations, which diminish as the number $M$ of open channels grows
\cite{Fyodorov1997,Savin1997}. In the limiting case $M\to\infty$ and
$T_{c}\to0$, all the resonances acquire just the same escape width (in units
of $t_{H}^{-1}$) $\sum_cT_{c}$ , which is often called the Weisskopf's width
\cite{Blatt}, so that the total width is $\gamma_T=\sum_cT_{c}+\gamma$. Then
further simplifications occur: $P_{ab}(\tau)=e^{-\gamma_T\tau}$ and
$J_{ab}(\tau)=[1-b_{2,\beta}(\tau)]e^{-\gamma_T\tau}$, that results finally
in
\begin{equation}\label{ScorrMT}
\fl C^{abcd}_S(\omega) =
\frac{(\delta^{ac}\delta^{bd}+\delta_{1\beta}\delta^{ad}\delta^{bc})T_aT_b
}{ \gamma_T-2\pi\rmi\omega } +
\delta^{ab}\delta^{cd}T_aT_c\int_{0}^{\infty}\!\rmd\tau
[1-b_{2,\beta}(\tau)]e^{-(\gamma_T-2\pi\rmi\omega)\tau}.
\end{equation}
For the case of $\beta=1$ this result (at zero absorption) was obtained
earlier by Verbaarschot \cite{Verbaarschot1986}. In the limit considered,
expression (\ref{ScorrMT}) is very similar to (\ref{ZcorrFTa}),
(\ref{ZcorrFTb}), so that the enhancement factor $W_{S,\beta}$ is given by
the same (\ref{W_Z}) where $\gamma$ is to be substituted with $\gamma_{T}$,
see Fig.~1(b) for an illustration. At $\gamma_T\gg1$ (large resonance
overlapping or strong absorption, or both) the dominating term in
(\ref{ScorrMT}) is the first one, which is known as the Hauser-Feshbash
relation \cite{Hauser1952}, see \cite{Moldauer1975a,Agassi1975,Friedman1985}
for discussion. Then $W_{S,\beta}=2/\beta=W_{Z,\beta}$ that can be also
understood as the consequence of the gaussian statistics of $S$ (as well as
of $Z$) in the limit of strong absorption \cite{Friedman1985}.

\section{Conclusions}

For open wave chaotic systems with preserved or broken TRS we have
calculated exactly the energy correlation function of impedance matrix
elements at arbitrary absorption and coupling. This function is found to be
related to the two-level cluster function, or to its form factor in the time
domain. The overall exponential decay due to uniform absorption is shown to
be the generic feature of any correlation function reduced to a two-point
spectral (resolvent) correlator, that follows simply from analytic
properties of the latter in the complex energy plane. The elastic
enhancement factor defined though the ratio of variances in reflection to
that in transmission diminishes gradually from the value $1+2/\beta$ at weak
absorption to $2/\beta$ at strong absorption.

The similar exact calculation for $S$-matrix elements has been performed in
the case of broken TRS, thus completing the well-known result
\cite{Verbaarschot1985} of preserved TRS. The corresponding enhancement
factor never reaches the maximum value $1+2/\beta$ at any finite resonance
overlapping. It attains the value $2/\beta$ in the limit of strong
absorption (independent of coupling) or at perfect coupling (independent of
absorption).
\\

We thank S. Anlage, U. Kuhl and H.-J. St\"ockmann for useful discussions.
The financial support
 by the SFB/TR 12 der DFG (D.V.S. and H.-J.S.) and
EPSRC grant EP/C515056/1
(Y.V.F.) is acknowledged.
\\


\begin{thebibliography}{10}

\bibitem{Stoeckmann}
St{\"{o}}ckmann H~J 1999 \emph{Quantum Chaos: An Introduction} (Cambridge,
UK:  Cambridge University Press).

\bibitem{Beenakker1997}
Beenakker C~W~J 1997 \emph{Rev. Mod. Phys.} \textbf{69} 731.

\bibitem{Alhassid2000}
Alhassid Y 2000 \emph{Rev. Mod. Phys.} \textbf{72} 895.

\bibitem{Doron1990}
Doron E, Smilansky U and Frenkel A 1990 \emph{Phys. Rev. Lett.} \textbf{65}
  3072.

\bibitem{Lewenkopf1992i}
Lewenkopf C~H, M{\"u}ller A and Doron E 1992 \emph{Phys. Rev. A} \textbf{45}
  2635.

\bibitem{Schaefer2003}
Sch{\"{a}}fer R, Gorin T, Seligman T~H and St{\"o}ckmann H~J 2003 \emph{J.
  Phys. A: Math. Gen.} \textbf{36} 3289.

\bibitem{Mendez-Sanchez2003}
{M\'endez-S\'anchez} R~A, Kuhl U, Barth M, Lewenkopf C~H and St{\"o}ckmann
H~J
  2003 \emph{Phys. Rev. Lett.} \textbf{91} 174102.

\bibitem{Schanze2005}
Schanze H, St{\"{o}}ckmann H~J, Mart\'{i}nez-{M}ares M and Lewenkopf C~H
2005
  \emph{Phys. Rev. E} \textbf{71} 016223.

\bibitem{Kuhl2005}
Kuhl U, Mart\'{i}nez-Mares M, M\'{e}ndez-S\'{a}nchez R~A and St{\"{o}}ckmann
  H~J 2005 \emph{Phys. Rev. Lett.} \textbf{94} 144101.

\bibitem{Barthelemy2005a}
Barth{\'{e}}lemy J, Legrand O and Mortessagne F 2005 \emph{Phys. Rev. E}
  \textbf{71} 016205.

\bibitem{Lobkis2003}
Lobkis O~I, Rozhkov I~S and Weaver R~L 2003 \emph{Phys. Rev. Lett.}
\textbf{91}
  194101.

\bibitem{Hul2004}
Hul O, Bauch S, Pakonski P, Savytskyy N, Zyczkowski K and Sirko L 2004
  \emph{Phys. Rev. E} \textbf{69} 056205.

\bibitem{Kogan2000}
Kogan E, Mello P~A and Liqun H 2000 \emph{Phys. Rev. E} \textbf{61} R17.

\bibitem{Beenakker2001}
Beenakker C~W~J and Brouwer P~W 2001 \emph{Physica E} \textbf{9} 463.

\bibitem{Savin2003i}
Savin D~V and Sommers H~J 2003 \emph{Phys. Rev. E} \textbf{68} 036211.

\bibitem{Fyodorov2003i}
Fyodorov Y~V 2003 \emph{JETP Lett.} \textbf{78} 250.

\bibitem{Savin2004}
Savin D~V and Sommers H~J 2004 \emph{Phys. Rev. E} \textbf{69} 035201(R).

\bibitem{Rozhkov2003}
Rozhkov I, Fyodorov Y~V and Weaver R~L 2003 \emph{Phys. Rev. E} \textbf{68}
  016204.

\bibitem{Rozhkov2004}
Rozhkov I, Fyodorov Y~V and Weaver R~L 2004 \emph{Phys. Rev. E} \textbf{69}
  036206.

\bibitem{Hemmady2005}
Hemmady S, Zheng X, Ott E, Antonsen T~M and Anlage S~M 2005 \emph{Phys. Rev.
  Lett.} \textbf{94} 014102.

\bibitem{Hemmady2005i}
Hemmady S, Zheng X, Antonsen T~M, Ott E and Anlage S~M 2005 \emph{Phys. Rev.
E}
  \textbf{71} 056215.

\bibitem{Zheng2004a}
Zheng X, Antonsen T~M and Ott E 2004 \emph{e-print}  cond--mat/0408327.

\bibitem{Zheng2004b}
Zheng X, Antonsen T~M and Ott E 2004 \emph{e-print}  cond--mat/0408317.

\bibitem{Mirlin2000}
Mirlin A~D 2000 \emph{Phys. Rep.} \textbf{326} 259.

\bibitem{Fyodorov1998i}
Fyodorov Y~V and Alhassid Y 1998 \emph{Phys. Rev. A} \textbf{58} R3375.

\bibitem{Fyodorov2004ii}
Fyodorov Y~V and Savin D~V 2004 \emph{JETP Lett.} \textbf{80} 725.
\newblock [Pis{'}ma v ZhETP \textbf{80}, 855 (2004)].

\bibitem{Savin2005}
Savin D~V, Sommers H~J and Fyodorov Y~V 2005 \emph{e-print}
  cond--mat/0502359.

\bibitem{Mahaux}
Mahaux C and Weidenm{\"u}ller H~A 1969 \emph{Shell-model Approach to Nuclear
  Reactions} (Amsterdam: North-Holland).

\bibitem{Verbaarschot1985}
Verbaarschot J~J~M, Weidenm{\"{u}}ller H~A and Zirnbauer M~R 1985
\emph{Phys.
  Rep.} \textbf{129} 367.

\bibitem{Sokolov1989}
Sokolov V~V and Zelevinsky V~G 1989 \emph{Nucl. Phys. A} \textbf{504} 562.

\bibitem{Lewenkopf1991}
Lewenkopf C~H and Weidenm{\"{u}}ller H~A 1991 \emph{Ann. Phys. (N.Y.)}
  \textbf{212} 53.

\bibitem{Fyodorov1997}
Fyodorov Y~V and Sommers H~J 1997 \emph{J. Math. Phys.} \textbf{38} 1918.

\bibitem{Nussenzveig}
Nussenzveig H~M 1972 \emph{Causality and Dispersion Relations} (New York:
  Academic Press).

\bibitem{Mehta2}
Mehta M~L 1991 \emph{Random Matrices}.
\newblock 2nd edn. (New York: Academic Press).

\bibitem{Guhr1998}
Guhr T, {M\"{u}ller-Groeling} A and Weidenm{\"{u}}ller H~A 1998 \emph{Phys.
  Rep.} \textbf{299} 189.

\bibitem{Lehmann1995a}
Lehmann N, Saher D, Sokolov V~V and Sommers H~J 1995 \emph{Nucl. Phys. A}
  \textbf{582} 223.

\bibitem{Lehmann1995b}
Lehmann N, Savin D~V, Sokolov V~V and Sommers H~J 1995 \emph{Physica D}
  \textbf{86} 572.

\bibitem{Brouwer1997ii}
Brouwer P~W and Beenakker C~W~J 1997 \emph{Phys. Rev. B} \textbf{55} 4695.

\bibitem{Sokolov2005}
Sokolov V~V 2004 \emph{e-print}  cond--mat/0409690.

\bibitem{Davy1987}
Davy J~L 1987 \emph{J. Sound Vib.} \textbf{115} 145.

\bibitem{Lobkis2000}
Lobkis O~I, Weaver R~L and Rozhkov I~S 2000 \emph{J. Sound Vib.}
\textbf{237} 281.

\bibitem{Alhassid1992}
Alhassid Y and Levine R~D 1992 \emph{Phys. Rev. A} \textbf{46} 4650.

\bibitem{Alhassid1998}
Alhassid Y and Fyodorov Y~V 1998 \emph{J. Phys. Chem. A} \textbf{102} 9577.

\bibitem{Zheng2005}
Zheng X, Hemmady S, Antonsen T~M, Anlage S~M and Ott E 2005 \emph{e-print}
  cond--mat/0504196.

\bibitem{Hemmady2005ii}
Hemmady S, Zheng X,  Antonsen T~M, Ott E and Anlage S~M 2005 \emph{e-print}
  nlin.CD/0506025; to be published in Acta Phys. Pol. A (Proceedings of the 2nd
Workshop on Quantum Chaos and Localization Phenomena, May 19--22, 2005,
Warsaw).

\bibitem{Verbaarschot1986}
Verbaarschot J~J~M 1986 \emph{Ann. Phys. (N.Y.)} \textbf{168} 368.

\bibitem{Gorin2002}
Gorin T and Seligman T~H 2002 \emph{Phys. Rev. E} \textbf{65} 026214.

\bibitem{Efetov1983}
Efetov K~B 1983 \emph{Adv. Phys.} \textbf{32} 53.

\bibitem{Pluhar1995}
Pluha{\v{r}} Z, Weidenm{\"{u}}ller H~A, Zuk J~A, Lewenkopf C~H and Wegner
F~J
  1995 \emph{Ann. Phys. (N.Y.)} \textbf{243} 1.

\bibitem{Efetov}
Efetov K~B 1996 \emph{Supersymmetry in Disorder and Chaos} (Cambridge, UK:
  Cambridge University Press).

\bibitem{Zuk1994}
Zuk J~A 1994 \emph{e-print}  cond--mat/0412060.

\bibitem{Fyodorov1995}
Fyodorov Y~V 1995 In E~Akkermans, G~Montambaux, J~L Pichard and
  J~Zinn-{J}ustin, eds., \emph{Mesoscopic Quantum Physics}, Proceedings of the
  {L}es-{H}ouches Summer School, Session LXI, p. 493 (Elsevier).

\bibitem{Dittes2000}
Dittes F~M 2000 \emph{Phys. Rep.} \textbf{339} 215.

\bibitem{Savin1997}
Savin D~V and Sokolov V~V 1997 \emph{Phys. Rev. E} \textbf{56} R4911.

\bibitem{Blatt}
Blatt J~M and Weisskopf V~F 1979 \emph{Theoretical Nuclear Physics} (Berlin:
  Springer-Verlag).

\bibitem{Hauser1952}
Hauser W and Feshbach H 1952 \emph{Phys. Rev.} \textbf{87} 366.

\bibitem{Moldauer1975a}
Moldauer P~A 1975 \emph{Phys. Rev. C} \textbf{11} 426.

\bibitem{Agassi1975}
Agassi D, Weidenm{\"{u}}ller H~A and Mantzouranis G 1975 \emph{Phys. Rep.}
  \textbf{22} 146.

\bibitem{Friedman1985}
Friedman W~A and Mello P~A 1985 \emph{Ann. Phys. (N.Y.)} \textbf{161} 276.

\end{thebibliography}

\end{document}